# Development of Self-Shunted Josephson Junctions For a Ten-Superconductor-Layer Fabrication Process: Nb/NbN$_x$/Nb Junctions

Sergey K. Tolpygo, *Senior Member, IEEE*, Ravi Rastogi, Terence Weir, Evan B. Golden, and Vladimir Bolkhovsky

*Abstract*—To increase integration scale of superconductor electronics, we are developing a new node, titled SFQ7ee, of the fabrication process at MIT Lincoln Laboratory. In comparison to the existing SFQ5ee node, we increased the number of fully planarized superconducting layers to ten and utilized NbN and NbN/Nb kinetic inductors to increase the possible inductor number density above a hundred million inductors per cm$^2$. Increasing the Josephson junction number density to the same level requires implementing self-shunted high-$J_c$ Josephson junctions. We investigated properties of Nb/NbN$_x$/Nb trilayer junctions as a potential replacement of high-$J_c$ Nb/Al-AlO$_x$/Nb junctions, where NbN$_x$ is a disordered, overstoichiometric, nonsuperconducting nitride deposited by reactive sputtering. Dependences of the $I_cR_n$ product and Josephson critical current density, $J_c$ on the NbN$_x$ barrier thickness and on temperature were studied in the thickness range from 5 nm to 20 nm. The fabricated junctions can be well described by the microscopic theory of SNS junctions, assuming no suppression of the energy gap in Nb electrodes near the NbN$_x$ interfaces and a Cooper pair decay length in the NbN$_x$ barrier, $\xi_n(T_c)$, of about 2.3 nm. Current-voltage characteristics of the junctions are well described by the resistively and capacitively shunted junction (RCSJ) model without excess current. In the studied range $J_c < 10$ mA/µm$^2$, the Nb/NbN$_x$/Nb junctions have lower values of the specific resistance $R_nA$, lower $I_cR_n$ products, and a stronger dependence of the $I_cR_n$ on temperature than the self-shunted or critically damped externally shunted Nb/Al-AlO$_x$/Nb junctions with the same critical current density; $A$ is the junction area, $I_c$ the junction critical current, and $R_n$ the effective shunting resistance.

## I. INTRODUCTION

SUPERCONDUCTOR electronics (SCE) hold the record of the highest clock frequency and of the lowest energy dissipation among all known electronics devices [1] - [7], though the ultralow levels of energy dissipation and the ultrahigh operating frequencies cannot be simultaneously achieved in the same digital circuits. The main problem of SCE is in a relatively low scale of integration [8], [9], *e.g.*, measured as device number density or the total device count per chip, resulting in a limited functionality compared to semiconductor (CMOS) circuits. For instance, the most advanced CMOS circuits use 3-nm and 5-nm manufacturing processes, whereas the most advanced fabrication processes for superconductor electronics is a 150-nm-linewidth process developed at MIT Lincoln Laboratory (MIT LL) [10]. Consequently, the currently achieved device number density in SCE (~ 1.5·10$^7$ Josephson junctions per cm$^2$) is a factor of 1000 lower than the transistor number density in CMOS circuits.

The problem of SCE integration scale also comes from the insufficiently developed circuit design tools. For instance, the SFQ cell library developed recently under the IARPA Supertools program [11] has device number density of only 5·10$^5$ Josephson junctions (JJs) per cm$^2$, i.e., a factor of 30 lower than what the fabrication technology is capable to provide [12], [13]. The existing cell libraries of superconductor adiabatic circuits based on Adiabatic Quantum Flux Parametrons (AQFP) provide similarly low device number densities: about 4.8·10$^5$ JJs/cm$^2$ [14], [15] for the SFQ5ee fabrication process at MIT LL [16], [17]; and about 1.6·10$^5$ JJs/cm$^2$ [18], [19] for the fabrication process developed at AIST, Japan. These device densities are almost five orders of magnitude lower than those in CMOS circuits.

The integration scale of SCE is set by the balance between the number densities of Josephson junctions and circuit inductors. Due to a low magnetic field penetration depth in niobium films, $\lambda_{Nb} \approx 90$ nm, Nb inductors are mainly geometrical (magnetic) in nature, their kinetic inductance is small for all practical widths of the inductor traces, $w$, and film thicknesses, $t$, when $w, t \gtrsim \lambda_{Nb}$. To increase the number density of inductors, we implemented kinetic inductors [20], [21] – materials with significantly larger kinetic inductance per unit length than Nb inductors due to $\lambda \gg \lambda_{Nb}$, e.g., Mo$_2$N, NbN, or NbTiN. This allows for a substantial reduction of the length, and hence the area, of the typical inductors without going to a sub-90-nm patterning. A cross section of the new process, titled SFQ7ee, is shown in Fig. 1, see also [22].

This material is based upon work supported by the Under Secretary of Defense for Research and Engineering under U.S. Air Force Contract No. FA8702-15-D-0001. Distribution statement A. Approved for public release: distribution is unlimited. *(Corresponding author: Sergey K. Tolpygo).*

All the authors are with the Lincoln Laboratory, Massachusetts Institute of Technology, Lexington, MA 02421 USA (emails: sergey.tolpygo@ll.mit.edu; ravi.rastogi@ll.mit.edu; weir@ll.mit.edu; evan.golden@ll.mit.edu; bolkv@ll.mit.edu).

Color versions of one or more of the figures in this article are available online at http://ieeexplore.ieee.org

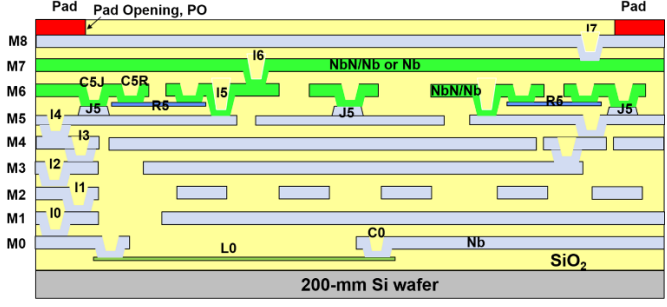

**Fig. 1.** Cross section of the SFQ7ee process [22]. Kinetic inductance layers M6 and M7 are deposited either as 200-nm-thick NbN or as NbN/Nb bilayer.

To increase the number density of Josephson junctions, $n_J$, we need to decrease their area, $A$, and proportionally increase the Josephson critical current density, $J_c$, because the critical current of the JJs, $I_c = J_c A$, is fixed by the circuit design. For the circular self-shunted JJs, the number density is

$$n_J = \chi\left(\left(\frac{4I_c}{\pi J_c}\right)^{\frac{1}{2}} + 2\delta\right)^{-2}, \quad (1)$$

where $\delta \sim 0.1$ µm is the required surround of the trilayer-type JJs by the base electrode and $\chi$ is the area filling factor, typically ~0.5. At the currently used $J_c = 0.1$ mA/µm² in the standard SFQ5ee process at MIT Lincoln Laboratory [16], [17], the $n_J$ can theoretically reach about 30M JJ/cm² at $I_c = 100$ µA and about 90M JJ/cm² at $I_c = 25$ µA; see Fig. 2. However, the currently used Nb/Al-AlO$_x$/Nb JJs are hysteretic at this critical current density and require external shunt resistors. The typical area of the externally shunted JJs is about 10x larger and the number density is ~10x smaller than those for the self-shunted JJs given by (1).

Nb/Al-AlO$_x$/Nb junctions become sufficiently self-shunted at $J_c \gtrsim 0.6$ mA/µm² [23], [24]. At this $J_c$, the junction number density can theoretically exceed 200M JJ/cm²; see the top open circles in Fig. 2. This requires reducing the minimum junction diameter $d$ from 0.6 µm in the existing SFQ5ee process down to about 0.25 µm. It is unknown however if statistical variation of such small junctions having a very thin AlO$_x$ barrier can be kept low enough for yielding very large-scale integrated circuits with a hundred million of JJs. Therefore, there is interest in developing alternative high-$J_c$ Josephson junctions with thicker barriers, e.g., of the SNS type where N is a highly disordered normal metal, or a doped (amorphous) semiconductor [25]-[30], or a planar bridge [31] and references therein. The hope is that these junctions may have lower parameter spreads than the high-$J_c$ SIS junctions due to much thicker barriers. The drawback of this approach is that all demonstrated types of SNS junctions have significantly lower $I_c R_n$ products, and hence lower operating frequency, than the self-shunted Nb/Al-AlO$_x$/Nb junctions [21], [23]; here $R_n$ is the effective shunting resistance of the junction.

In parallel with the development of the SFQ7ee process with self-shunted Nb/Al-AlO$_x$/Nb JJs of $J_c = 0.6$ mA/µm², we investigate alternative self-shunted junctions for a potential use in future very large-scale integration (VLSI) process nodes. In this work we investigated properties of Nb/NbN$_x$/Nb Josephson junctions as a vehicle for optimizing a nonsuperconducting NbN$_x$ (x>1) barrier for all-niobium-nitride NbN/NbN$_x$/NbN junctions. The use of NbN-based junctions in the SFQ7ee process is more desirable than of Nb-based ones because the junctions are wired by the NbN layer M6, see Fig. 1.

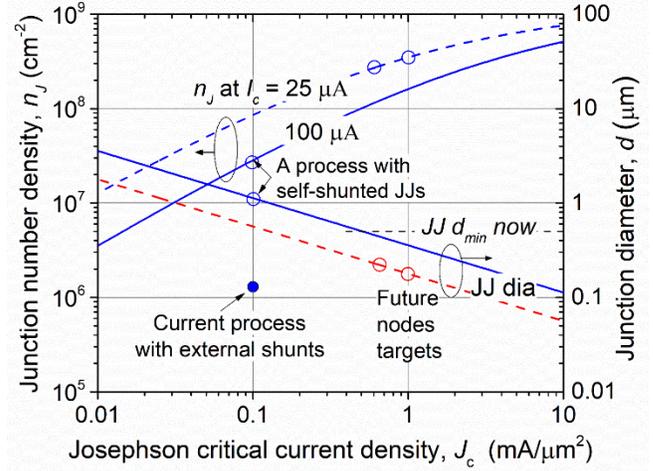

**Fig. 2.** Josephson junction number density as a function of the process Josephson critical current density (left axis scale) for two values of the typical critical current: 25 µA (the top dashed blue curve) and 100 µA (solid blue curve). The junction diameter required to achieve these critical currents is shown by the right axis scale for $I_c = 25$ µA (bottom red dashed curve) and 100 µA (solid curve). The currently used Nb/Al-AlO$_x$/Nb junctions have $J_c = 0.1$ mA/µm²; (●) – the demonstrated circuit density in the SFQ5ee process with externally shunted JJs; (○) theoretical value (1) for $J_c = 0.1$, 0.6, and 1 mA/µm².

## II. FABRICATION OF NB/NBN$_x$/NB TRILAYER JUNCTIONS

A truncated version of the SFQ5ee fabrication process [17], [32], starting from the layer M4 on, was used. Nb/NbN$_x$/Nb trilayers were deposited on 200-mm wafers having the patterned Nb layer M4 covered by the planarized SiO$_2$ interlayer dielectric. The trilayers were deposited in an Endura PVD (Applied Materials) cluster tool. Deposition parameters of 150-nm-thick Nb electrodes were identical to the ones used for Nb/AlO-AlO$_x$/Nb junctions in the standard process [32]. Reactive sputtering of Nb in N$_2$+Ar flowing gas mixture was used for depositing NbN$_x$ barrier with thicknesses in the range from 5 nm to 20 nm, at room temperature.

Dependences of the resistivity, $\rho$, and superconducting critical temperature, $T_c$ of the reactively sputtered NbN$_x$ films on the deposition parameters in the Endura PVD system were studied in [22]. It was observed that the resistivity of NbN$_x$ films monotonically increases with increasing N$_2$ partial pressure, whereas superconducting films, with the highest $T_c$ of 16 K, are only produced in a relatively narrow range of N$_2$ partial pressures; see [22, Fig. 5]. Highly nonstoichiometric and nonsuperconducting films are produced at lower and higher N$_2$ contents in the sputtering gas mixture than the optimal range. Since we need high-resistivity nonsuperconducting barriers for the SNS-type junctions, we used much higher N$_2$ to Ar flow ratios for depositing the NbN$_x$ barriers than for depositing the stoichiometric NbN films: N$_2$/(N$_2$+Ar) = 0.818 and 0.905.

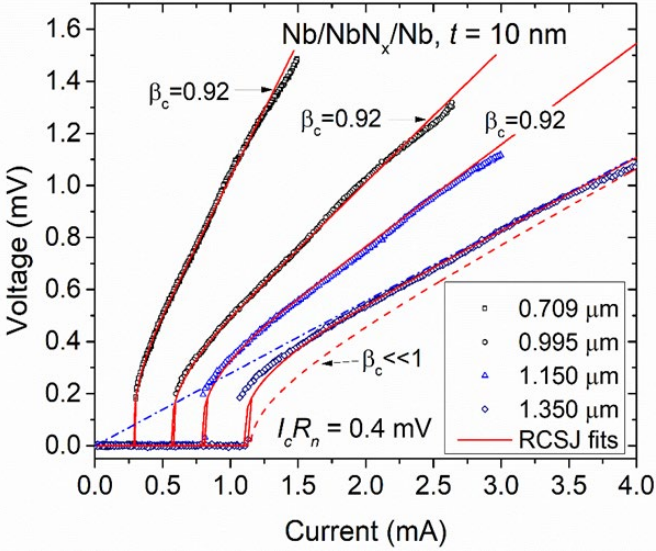

**Fig. 3.** The typical current-voltage characteristic of Nb/NbN$_x$/Nb Josephson junctions at 4.2 K; the junctions are from 0.709 μm to 1.35 μm in diameter, from left to right. Solid curves are fits to the RCSJ model giving $I_cR_n = 0.4$ mV at the barrier thickness of 10 nm. The barrier was deposited using N$_2$/(N$_2$+Ar) =0.905.

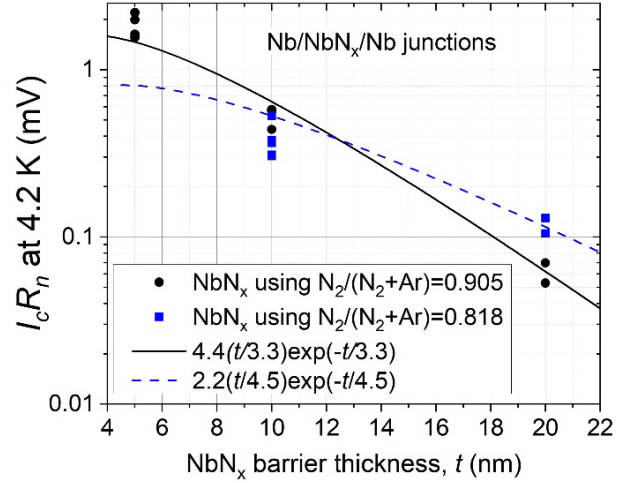

**Fig. 4.** Dependences of the $I_cR_n$ product of Nb/NbNx/Nb Josephson junctions at 4.2 K on the NbN$_x$ barrier thickness, $t$, for the barriers deposited at two different nitrogen contents in the sputtering gas mixture. Solid and dashed curves are functions $I_cR_n = V_0 \frac{t}{\xi_n(T)} \exp\left(-\frac{t}{\xi_n(T)}\right)$ [34]. [35] with parameters $V_0$ and $\xi_n$(4.2 K) given in the legend in units of mV and nm, respectively.

The trilayers were patterned using a 248-nm photography (a Canon FPA-3000 EX4 stepper) and a high-density plasma etching in a Cl$_2$/BCl$_3$/Ar mixture (in an Applied Materials Centura etch cluster). Optical emission end-point detection was used to stop etching the top Nb electrode on the NbN$_x$ barrier. Due to a much lower reactivity of NbN$_x$ in comparison to Al, implementation of NbN$_x$ barriers instead of Al-AlO$_x$ allowed for the elimination of the junction mesa anodization step following right after the etching of the junction top electrode of Nb/Al-AlO$_x$/Nb junctions.

After pattering the top and bottom electrodes of the junctions, the wafer was planarized by depositing a thick layer of SiO$_2$ and using chemical-mechanical polishing to reach the required interlayer dielectric thickness. Vias to the junctions' top and bottom electrodes were etched and filled in by depositing a 200-nm Nb layer M6. These steps of the fabrication are identical to the standard SFQ5ee process described in [16], [17], [32].

Sufficient internal damping of Nb/NbN$_x$/Nb junctions, see Sec. III, allowed us to eliminate multiple fabrication steps related to making external shunt resistors: resistor film deposition, resistor film patterning, dielectric layer deposition above the patterned resistors and its planarization, photolithography of and etching vias to the resistors; see these steps in [8, Fig. 5] and [32, Fig. 10].

III. PROPERTIES OF NB/NBN$_X$/NB JUNCTIONS

As an example, the typical current-voltage (I-V) characteristics of the fabricated Nb/NbN$_x$/Nb junctions with the barrier thickness $t$=10 nm are shown in Fig. 3 for the circular junctions with various design diameters. The I-V characteristics are nearly nonhysteretic and can be fitted well by the resistively and capacitively shunted junction (RCSJ) model with the McCumber-Stewart parameter $\beta_c = 2\pi I_c R_n^2 C/\Phi_0$ in the range from about 0.9 to 1 as shown in Fig. 3 by the solid red curves; here $C$ is the junction capacitance. The fit to the RSJ model ($C = 0$) is also shown by the dashed curve as well the ohmic dependence $V = IR_n$ (dash-dot line). It is seen from the linear fit that there is no excess current which is usually observed in SNS junctions and SIS junctions with high barrier transparency; see e.g. [20, Fig. 12], [23, Fig. 9], [24, Fig. 2], and [33, Fig. 2].

Dependence of the $I_cR_n$ product on the barrier thickness is shown in Fig. 4 for the barriers deposited at two different nitrogen contents in the sputtering gas mixture. The scattering of the data points reflects the variation of the $I_cR_n$ across the wafer and from wafer to wafer. This variation is caused by changes in the thickness of NbN$_x$ films produced by reactive sputtering. The NbN$_x$ film cross-wafer thickness distribution was studied in [21, Fig. 9]. It has a doughnut-like shape (lower in the wafer center and around the edges and higher in the wafer middle) reflecting the magnetic field distribution in a magnetron sputtering gun.

Assuming the simplest exponential dependence [34], [35]

$$I_cR_n = V_0 \frac{t}{\xi_{n(T)}} \exp\left(-\frac{t}{\xi_{n(T)}}\right), \qquad (2)$$

we can estimate the Cooper pair decay length in the barrier, $\xi_n$(at 4.2 K), as about 3.3 nm and 4.5 nm for the barriers deposited, respectively, in N$_2$/(N$_2$+Ar)=0.905 and 0.818. These values of the $\xi_n$ are comparable to the coherence length observed in junctions with amorphous Si and doped-Si barriers, or with TiN$_x$ barriers; see [25] – [30], [33], [36]. Note that (2) is valid only at $t \gg \xi_n$, which is the case only for the largest barrier thicknesses used; more general treatment is given in IV.A.

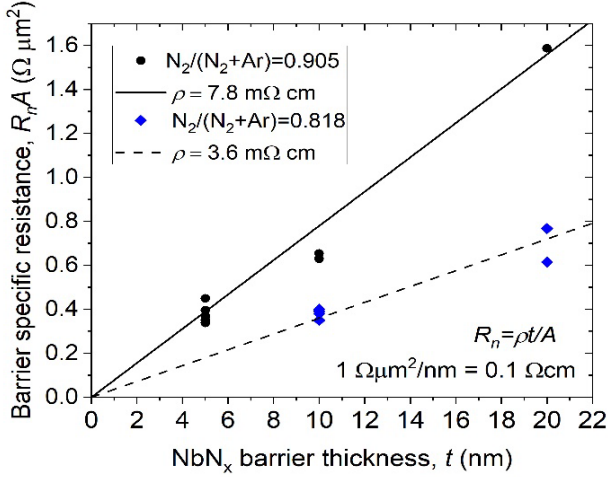

**Fig. 5.** Dependence of the NbN$_x$ barrier specific resistance $R_nA$ of Nb/NbN$_x$/Nb Josephson junctions on the barrier thickness for the barriers deposited at two different nitrogen contents in the sputtering gas mixture. Solid and dashed lines correspond, respectively, to the barrier material resistivity of 7.8 and 3.6 mΩ·cm; the lines slope conversion is 1 Ω·μm² nm$^{-1}$ = 0.1 Ω·cm.

The junction specific resistance $R_nA$ determined from the I-V characteristics is shown in Fig. 5. In general, we expect $R_nA = \rho_n t + 2R_b$, where $R_b$ is the interface resistance between Nb and NbN$_x$ per unit area. The linear fits shown in Fig. 5 are at $R_b = 0$. The inferred NbN$_x$ barrier resistivity $\rho$ strongly depends on the nitrogen content and is a thousand times higher than residual resistivity of the Nb electrodes $\rho_{Nb} \approx 4$ μΩ·cm.

Dependence of the Josephson critical current density, $J_c$ at 4.2 K on the barrier thickness is shown in Fig. 6. If (2) is valid and the interface resistance is small, we can expect that $J_c$ changes as

$$J_c = \frac{V_0}{\rho \xi_n(T)} \exp\left(-\frac{t}{\xi_n(T)}\right) \quad (3)$$

with the same coherence length as in (2). Dependence (3) is shown in Fig. 6 by the solid and dashed lines corresponding to the junction parameters inferred from Fig. 4 and Fig. 5: $V_0 = 4.4$ mV, $\xi_n(4.2K) = 3.3$ nm, the barrier resistivity $\rho = 0.078$ Ω·μm for the NbN$_x$ barriers deposited using N$_2$/(N$_2$+Ar)=0.905 (solid line); $V_0 = 2.2$ mV, $\xi_n(4.2K) = 4.5$ nm, $\rho = 0.036$ Ω·μm for the barriers deposited using N$_2$/(N$_2$+Ar)=0.818 (dashed line).

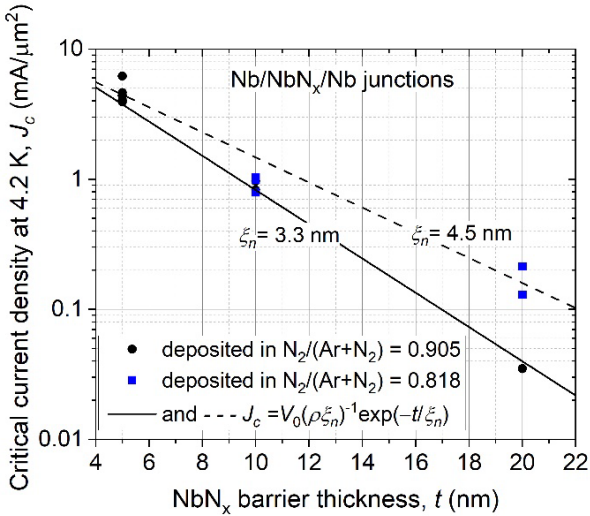

**Fig. 6.** Dependence of the Josephson critical current density at 4.2 K of Nb/NbN$_x$/Nb Josephson junctions on the barrier thickness for the NbN$_x$ barriers deposited at two different nitrogen contents in the sputtering gas mixture; the same wafers and the deposition conditions as in Fig. 4. Solid and dashed lines are dependences (3) with the same parameters $V_0$ and $\xi_n(4.2K)$ as in Fig. 4 and the barrier resistivity values given in Fig. 5.

Temperature dependence of the $I_cR_n$ product of Nb/NbN$_x$/Nb junctions is shown in Fig. 7 for two barrier thicknesses. It represents the $I_c(T)$ dependence since the junction resistance $R_n$ is temperature independent in this range.

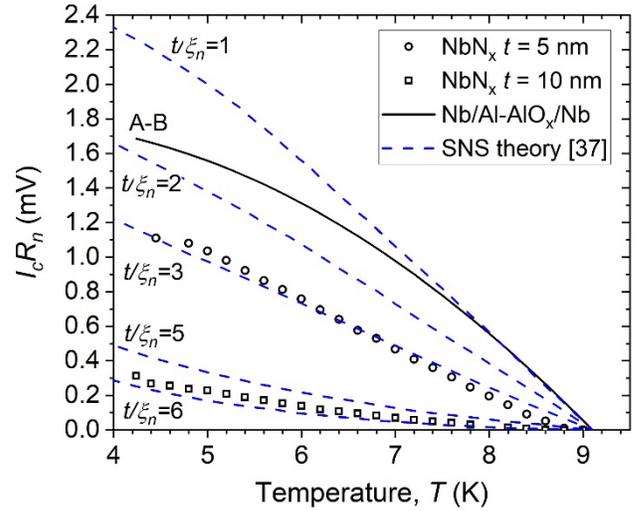

**Fig. 7.** The typical temperature dependences of the $I_c(T)R_n$ product of Nb/NbN$_x$/Nb Josephson junctions with different NbN$_x$ barrier thicknesses deposited using N$_2$/(N$_2$+Ar)=0.905 flow ratio: (○) – 5 nm; (□) – 10 nm. Dashed curves are theoretical dependences [37] for SNS junctions with "rigid" boundary conditions at different values of $t/\xi_n$, where $\xi_n \equiv \xi_n(T_c)$ is the Cooper pair decay length in the N-barrier at $T_c$ of the S electrodes. The normalized dependences obtained in [37] were rescaled using the zero-temperature energy gap in Nb $\Delta(0)/e = 1.445$ mV and $T_c = 9.1$ K. Solid curve, marked A-B, is the $I_c(T)R_n$ dependence of Nb/Al-AlO$_x$/Nb JJs in the SFQ5ee process, which coincides with the Ambegaokar-Baratoff expression (6) with $\alpha = 0.8$, the energy gap temperature dependence given by (7), and $\Delta(0)/e = 1.445$ mV; see IV.B.

## IV. DISCUSSION

### A. Coherence Length and $I_cR_n$ Temperature Dependence

Microscopic theories describing properties of SNS, SS'S, SINS, SNINS, etc. types of Josephson junctions with various types of barriers exist; see [37]-[40]. The results critically depend on several parameters describing the suppression of superconductivity in the junction electrodes due to the proximity effect and the current flowing through the junction, temperature, and the ratio $t/\xi_n$ of the barrier thickness $t$ to the Cooper pair decay length (coherence length) at $T = T_c$ of the junction electrodes, $\xi_n \equiv \xi_n(T_c)$. Coherence length in a disordered non-superconducting barrier is given by the well-known expression [34], [35]

$$\xi_n(T) = \left(\frac{\hbar D}{2\pi k_B T}\right)^{1/2}, \quad (4)$$

where $D$ is the carrier diffusion coefficient. The NbN$_x$ barriers have a small metallic conductivity in comparison to the Nb electrodes, $\rho_n \gg \rho_s$, and a low interface resistance with Nb. Hence, parameters $\gamma$, $\gamma_B$ and $\gamma_J$ describing the suppression of superconductivity in the electrodes due to the proximity effect and the current flowing through the junction [39], [40] are small $\gamma = \frac{\rho_s \xi_s}{\rho_n \xi_n} \sim 0.01$, $\gamma_B = R_B t/\xi_n \approx 0$, and $\gamma_J = \frac{\rho_s \xi_s}{\rho_n t} \sim 5 \cdot 10^{-3}$, where $\xi_s$ is the coherence length in the junction electrodes. In this case, the theory of SNS junctions based on the "rigid" boundary conditions, using the unperturbed value of the order parameter $\Delta$ in the S region, developed in [37], [38] applies; see also [40, Fig. 1].

Temperature dependences of the $I_c R_n$ product of SNS junctions [37] are shown in Fig. 7 along with the experimental data for the Nb/NbN$_x$/Nb junctions with $t = 5$ and 10 nm. The dashed curves represent the theoretical results [37] for several values of $t/\xi_n$, rescaled using the measured values of the energy gap in our Nb/Al-AlO$_x$/Nb junctions: $\Delta(4.2K)/e = 1.40$ mV and $\Delta(0)/e = 1.445$ mV. Close to the $T_c$, the $I_c R_n$ product is given by

$$I_c R_n = \frac{2\Delta(T)^2}{\pi e k_B T} \sum_{n\geq 0} \frac{1}{(2n+1)^2} \frac{l_n}{\sinh l_n}, \quad (5)$$

where $l_n = (2n+1)^{1/2} \frac{t}{\xi_n(T)}$ and $\Delta(T)$ is the order parameter at the S/N interface [37].

The measured $I_c R_n$ values and temperature dependences for the junctions with $t = 5$ nm and 10 nm are well described by the theory [37] at, respectively, $t/\xi_n \approx 3$ and $t/\xi_n \approx 6$. This suggests that $\xi_n \approx 1.7$ nm. Since $\xi_n(T) = \xi_n(T_c/T)^{1/2}$, the coherence length at a liquid helium (LHe) temperature is $\xi_n(4.2K) = 1.465\xi_n \approx 2.5$ nm, using niobium $T_c = 9.1$ K. The latter length is a bit smaller than the 3.3 nm value estimated using (2) and (3); see Figs. 4 and 5.

To compare the data in Fig. 4 with the SNS junction theory [37], $I_c R_n$ dependences on $T/T_c$ presented in [37] and [38] were recalculated to obtain the $I_c R_n(4.2K)$ dependence (at $T/T_c = 0.466$) on $t/\xi_n$ for Nb electrodes. This dependence is shown in Fig. 8 by the solid curve. To check the range of applicability of the large thickness limit (2) at $T=4.2$ K, (2) was fit to the full theory as shown in Fig. 8 by the dashed curve, using $t/\xi_n(T) = t/(1.465\xi_n)$ and $V_0$ as the only fitting parameter, giving $V_0 = 3.9$ mV. Clearly, (2) is a very good approximation to the full theory at $t/\xi_n \gtrsim 3$. At $t/\xi_n \to 0$, the SNS theory gives $I_c R_n(0) = \frac{1.32\pi\Delta(0)}{2e} = 3.0$ mV at $T = 0$, and $I_c R_n(4.2K) = 2.52$ mV (for the junctions with Nb electrodes) in agreement with the theory of the "dirty" point contacts [41].

Next, we rescaled the barrier thicknesses of the junctions in Fig. 4 using $\xi_n$ as the only fitting parameter to get the best description of the data by the SNS theory [37] (solid curve). This gave $\xi_n(T_c) = 2.33$ nm and 2.65 nm for the barriers deposited using N$_2$/(N$_2$+Ar)=0.905 and 0.818 flow ratios, respectively. These values of $\xi_n(9.1$ K$)$ translate into $\xi_n(4.2K) = 3.4$ nm and 3.9 nm in a good agreement with the values estimated from the fits to (2), shown in Fig. 4 and Fig. 5, which used $V_0$ and $\xi_n(4.2K)$ as fitting parameters.

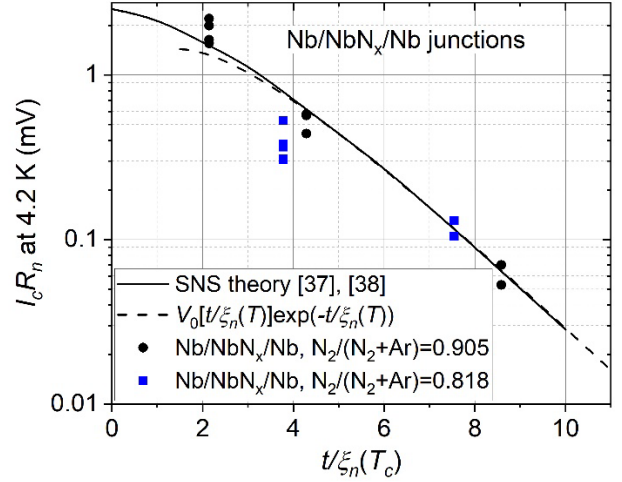

**Fig. 8.** Theoretical dependence of the $I_c R_n$ product of Nb-based SNS junctions at LHe temperature $T=4.24$ K on the $t/\xi_n(T_c)$ ratio (solid curve) based on the results in [37], [38]; the theory uses the rigid boundary conditions and assumes the absence of the S/N interface resistance. To rescale the normalized values given in [37] and [38], the zero-temperature gap $\Delta(0)/e$ of 1.445 mV was used. Dashed curve – the best fit of the theoretical dependence to (2), giving $V_0 = 3.9$ mV. Rescaling the barrier thicknesses of the junctions in Fig. 4 using $\xi_n(T_c)$ as a fitting parameter to get the best description by the theory (the solid curve) resulted in $\xi_n(T_c) = 2.33$ nm and 2.65 nm for the barriers deposited using N$_2$/(N$_2$+Ar)=0.905 and 0.818, respectively.

*B. Comparison With Nb/Al-AlOx/Nb SIS-Type Junctions*

For comparison, the typical $I_c(T)R_n$ dependence for Nb/Al-AlO$_x$/Nb junctions fabricated in the SFQ5ee process is shown in Fig. 7 by a solid curve marked A-B. This dependence is well known [42] to very closely follow the Ambegaokar-Baratoff expression [43]

$$I_c R_n = \alpha \frac{\pi\Delta(T)}{2e} \tanh\frac{\Delta(T)}{2k_B T}, \quad (6)$$

with the energy gap $\Delta(T)$ in Nb electrodes given by the strong-coupling expression [44]

$$\frac{\Delta(T)}{\Delta(0)} = \tanh\left[\frac{T_c \Delta(T)}{T \Delta(0)}\right]. \quad (7)$$

The numerical coefficient $\alpha$ accounts for the strong-coupling effects in the junction electrodes [45] and depends on the thickness of the Al interlayer [46]. For the SFQ5ee fabrication process using 8 nm Al layer, $\alpha \approx 0.8$. Deviations from (6) are very small even in Nb/Al-AlO$_x$/Nb junctions with very high Josephson critical current densities $J_c \approx 2.5$ mA/μm$^2$ [47], whose transport properties are determined by multiple Andreev reflections via atomic-size conducting channels with a broad distribution of transmission probabilities.

Near $T_c$, the unshunted Nb/Al-AlO$_x$/Nb junctions have the same $I_c(T)R_n$ values as can be expected from Nb/NbN$_x$/Nb and other types of Nb-based SNS junctions at $t/\xi_n =1$. However, for superconductor digital electronics applications, the relevant parameter is not the $I_c(T)R_n$ but the characteristic voltage $V_c = I_c(T)R_s$ of the damped junctions, where $R_s$ is the effective shunting resistance providing the critical damping $\beta_c =1$. The $V_c$ determines the characteristic frequency $f_c = V_c/\Phi_0 = (J_c/2\pi C_s\Phi_0)^{1/2}$ that corresponds to the maximum operating frequency of simple single flux quantum (SFQ) digital circuits, e.g., T flip-flops [48]; $C_s$ is the junction capacitance per unit area, $\Phi_0 \approx 2.07$ mV·ps the superconducting flux quantum. More complex, clocked, SFQ digital circuits typically operate only to about $0.15f_c$. For instance, the critically damped Nb/Al-AlO$_x$/Nb JJs with $J_c$=0.1 mA/µm$^2$ in the SFQ5ee fabrication process [17] have $V_c$ =0.67 mV and $f_c$ =324 GHz. The SFQ circuits produced in this process typically operate up to ~50 GHz; see, e.g. [49] – [51] and references therein.

*C. Drawbacks of Implementing SNS-Type Junctions*

From Fig. 6, the $J_c$=0.1 mA/µm$^2$ required for a straightforward one-to-one replacement of the externally shunted Nb/Al-AlO$_x$/Nb JJs in the SFQ5ee process by the self-shunted Nb/NbN$_x$/Nb junctions can be achieved at the barrier thickness of about 20 nm by slightly adjusting the nitrogen content between the two values used. Alternatively, the barrier thickness can be reduced to about 16 nm while keeping the flow ratio N$_2$/(N$_2$+Ar)=0.905. Unfortunately, according to Fig. 4, the $I_cR_n$ product at these parameters is expected to be only about 0.15 mV that is significantly lower than the $V_c$ =0.67 mV of the critically damped Nb/Al-AlO$_x$/Nb JJs [23]. So, implementing Nb/NbN$_x$/Nb junctions can give a significant increase in the circuit density by getting rid of the shunt resistors but will significantly reduce the maximum possible circuit speed, to about 15 GHz.

Self-shunted Nb/Al-AlO$_x$/Nb JJs with $J_c$ =0.6 mA/µm$^2$ have $V_c \approx$1.2 mV [23], a factor of 2 to 3 higher than the Nb/NbN$_x$/Nb junctions with the same $J_c$; see Figs. 4, 5. This difference is not as dramatic as for the lower-$J_c$ junctions, and Nb/NbN$_x$/Nb JJs could potentially be used in circuits not requiring the record high operating speeds. However, from the data in Figs. 3-5, Nb/NbN$_x$/Nb junctions with this $J_c$ have 2 to 3 times lower resistance $R_n$ or rather lower specific resistance $R_nA$~0.5 Ω µm$^2$ than the self-shunted Nb/Al-AlO$_x$/Nb JJs. Matching the lower-impedance junctions to passive transmission lines (PTLs) would require wider transmission lines that would go against increasing the scale of integration.

This PTL-JJ matching problem is quite severe even with Nb/Al-AlO$_x$/Nb JJs. In SFQ circuits, the typical Nb/Al-AlO$_x$/Nb JJs have $I_c$~0.1 mA, resistance $V_c/I_c$ ~ 7 Ω, and require about 6-µm-wide matching striplines for the data and clock transfer [52], [53]. Such wide PTLs are incompatible with the very large-scale integration requirements. Therefore, implementing any JJs with lower $R_n$ than the $R_s$ of the Nb/Al-AlO$_x$/Nb JJs, would make the PTL integration problem yet more severe.

Reaching comparable to Nb/Al-AlO$_x$/Nb $V_c$ values in Nb/NbN$_x$/Nb JJs or other types of Nb-based SNS junctions requires thin barriers with $t/\xi_n < 4$, according to Fig. 8. Unfortunately, these junctions will have very high values of $J_c$, and very low values of the specific resistance $R_nA$, which are respectively much higher and much lower than the $J_c$ and $R_nA$ values in Nb/Al-AlO$_x$/Nb JJs with the same value of $V_c$. Hence, fabrication of the JJs with critical currents used for superconductor digital electronics, $I_c = J_cA \lesssim$0.2 mA, would require much finer photolithography to define smaller junction areas and more stringent requirements to the area definition, to keep the $I_c$ spreads small, than those currently used for making Nb/Al-AlO$_x$/Nb junctions.

It is also very unlikely that any type of deposited barriers could have a better cross-wafer uniformity and wafer-to-wafer reproducibility than the AlO$_x$ tunnel barriers obtained by Al oxidation in an O$_2$-containing gas mixtures. This is because the oxide growth rate is mainly determined by the local O$_2$ pressure that is identical at all points on the wafer in the case of static oxidation. Contrary to this, the deposition rate at reactive sputtering or a plasma-enhanced chemical vapor deposition (PECVD) is determined by a cumulative effect of the flowing gas pressure distribution and the target material sputtering yield distribution (or plasma configuration in the PECVD case), both hardware-dependent. The former cannot be as uniform as in the static case and the latter contributes at least 2% standard deviation to the film thickness variation across 200-mm wafers. Hence, $J_c$ uniformity across the wafer and wafer-to-wafer reproducibility cannot be better than the thickness uniformity and thickness reproductivity of the deposited barrier film. For instance, the most recent data on Nb-based junction with deposited amorphous-Si barriers [54] show inferior properties and junction uniformity to Nb/Al-AlO$_x$/Nb JJs at the same values of $J_c$.

V. CONCLUSION

We investigated properties of Nb/NbN$_x$/Nb Josephson junctions to evaluate the potential of NbN$_x$ as a barrier for high-$J_c$ self-shunted Josephson junctions. We found that the junctions are well described by the RCSJ model and have reduced values of the $R_nA$ and $I_cR_n$ compared to Nb/Al-AlO$_x$/Nb junctions at the same $J_c$ values. High $I_cR_n$ products comparable to those in externally shunted and self-shunted Nb/Al-AlO$_x$/Nb JJs can be achieved using thin barriers, $t \lesssim 5$ nm, giving very high Josephson critical current densities at very low barrier specific resistances. Consequently, the integrated circuit fabrication process would require defining much smaller junction areas than is currently required for the self-shunted Nb/Al-AlO$_x$/Nb junctions while delivering the same relative spread of the critical currents. Temperature dependence of the critical current of Nb/NbN$_x$/Nb junctions in the typical range of operation of superconductor integrated circuits (4 K to 6 K) is stronger than of Nb/Al-AlO$_x$/Nb junctions, which may also complicate implementation of Nb/NbN$_x$/Nb JJs in large-scale electronic systems.

The mentioned drawbacks of Nb/NbN$_x$/Nb junctions are typical for all types of SNS and bridge-type Josephson junctions with various types of (poorly) conducting barriers, e.g., used in [25] – [31], [33], [36], [54]. As a result, they are more suitable for applications requiring JJs with large critical currents and small $I_cR_n$ products, such as various types of dc voltage standards, rather than for energy-efficient, high-speed digital integrated circuits requiring junctions with relatively small,

≲0.2 mA, critical currents and high $I_cR_n$ products; see also [55] and references therein.

## Acknowledgement

We are grateful to Justin Mallek and David Kim for their help on different stages of this project, and to Mark Gouker and Leonard Johnson for their interest in this research. Any opinions, findings, conclusions, or recommendations expressed in this material are those of the authors and do not necessarily reflect the views of the Under Secretary of Defense for Research and Engineering or the U.S. Government. Notwithstanding any copyright notice, U.S. Government rights in this article are defined by DFARS 252.227-7013 or DFARS 252.227-7014 as detailed above. Use of this article other than as specifically authorized by the U.S. Government may violate any copyrights that exist in this article. The U.S. Government is authorized to reproduce and distribute reprints for Governmental purposes notwithstanding any copyright annotation thereon.